\documentclass[aps,prb,preprint,groupedaddress,showpacs,amsmath,amssymb]{revtex4}

\usepackage{graphicx}
\usepackage{dcolumn}
\usepackage{bm}

\begin{document}

\title{High temperature ferromagnetism in GdFe$_2$Zn$_{20}$:  large, local moments embedded in the nearly ferromagnetic Fermi liquid compound YFe$_2$Zn$_{20}$.}

\author{S. Jia$^1$, S. L. Bud'ko$^1$, G.D. Samolyuk$^2$, P. C. Canfield$^1$}
\affiliation{$^1$Ames Laboratory US DOE and Department of Physics and Astronomy, Iowa State University, Ames, IA
50011\\ $^2$Ames Laboratory US DOE and Department of Chemistry, Iowa State University, Ames, IA 50011}

\date{\today}

\begin{abstract}

The RFe$_2$Zn$_{20}$ series manifests strongly correlated electron behavior for the non-magnetic R = Y member and
remarkably high temperature, ferromagnetic ordering ($T_C$ = 86 K) for the local moment bearing R = Gd member (a
compound that is less than 5\% atomic Gd).  In contrast, the isostructural RCo$_2$Zn$_{20}$ series manifests a
more typical ordering temperature ($T_N$ = 5.7 K for GdCo$_2$Zn$_{20}$) and YCo$_2$Zn$_{20}$ does not show signs
of correlated electron behavior. Studies of R(Fe$_x$Co$_{1-x}$)$_2$Zn$_{20}$ (R = Gd, Y), combined with
bandstructure calculations for the end members, reveal that YFe$_2$Zn$_{20}$ is a nearly ferromagnetic Fermi
liquid and that the remarkably high $T_C$ associated with GdFe$_2$Zn$_{20}$ is the result of submerging a large
local moment into such a highly polarizable matrix. These results indicate that the RFe$_2$Zn$_{20}$ series, and
more broadly the RT$_2$Zn$_{20}$ (T = Fe, Co, Ni, Mn, Ru, Rh, Os, Ir, Pt) isostructural family of compounds, offer
an exceptionally promising phase space for the study of the interaction between local moment and correlated
electron effects near the dilute R limit.

\end{abstract}



\maketitle

The field of condensed mater physics has been interested in the effects of electron correlations from it inception
\cite{mor85a,bro90a}.  To this day, the properties of elemental Fe as well as Pd continue to present problems that
interest both experimentalists as well as theorists \cite{zel04a}.  In particular materials such as Pd or Pt, that
are just under the Stoner limit (often referred to as nearly ferromagnetic Fermi liquids), or materials just over
the Stoner limit, such as ZrZn$_2$ or Sc$_3$In on the ferromagnetic side, are of particular interest
\cite{mor85a,bro90a,zel04a}.  Of even greater interest are new examples of nearly ferromagnetic Fermi liquids that
can be tuned with a greater degree of ease than the pure elements:  i.e. can accommodate controlled substitutions
on a number of unique crystallographic sites in a manner that allows for (i) a tuning of the band filling / Fermi
surface and (ii) the introduction of local moment bearing ions onto a unique crystallographic site.  Such a
versatile system would open the field to a greater range of experimental studies of strongly correlated electronic
states as well as potentially allow for more detailed studies of quantum criticality and possibly even novel
superconducting states.

In this letter we present our first results of an extensive study of the dilute, rare earth bearing, intermetallic
series RT$_2$Zn$_{20}$ (R = rare earth and T = transition metal) which supports a wide range of R ions for T in
and near the Fe, Co and Ni columns of the periodic table.  In particular, in this letter we will show how
YFe$_2$Zn$_{20}$ is an archetypical example of a nearly ferromagnetic Fermi liquid and how, by imbedding Gd ions
into this highly polarizable medium, GdFe$_2$Zn$_{20}$ can have a remarkably high ferromagnetic ordering
temperature of 86 K, even though it contains less than 5\% atomic Gd and the Fe is not moment bearing in the
paramagnetic state.

The RT$_2$Zn$_{20}$ series of compounds was discovered in 1997 by Nasch et al. \cite{nas97a} and is isostructural
to the cubic CeCr$_2$Al$_{20}$ structure ($Fd\bar{3}m$ space group)\cite{kri68a,thi98a}.  The rare earth and
transition metal ions each occupy their own single, unique crystallographic sites ($8a$ and $16d$ respectively)
whereas the Zn ions have three unique crystallographic sites ($96g$, $48f$ and $16c$). In addition, the rare earth
site is one of cubic point symmetry ($\bar{4}3m$).  The coordination polyhedra for R and T are comprised fully of
Zn, meaning that there are no R-R, T-T or R-T nearest neighbors and the shortest R-R spacing is $\sim 6$ \AA. The
lattice parameters for the series are $\sim 14$ \AA. Although the crystallography of this series was well
detailed, there have been, to date, no measurements of these compounds' physical properties. This, to some extent,
is not unexpected since the limited data sets available on the RT$_2$Al$_{20}$ compounds \cite{thi98a,moz98a}
indicated very low ordering temperatures, consistent with the very low R-concentrations.

Single crystals of RFe$_2$Zn$_{20}$, RCo$_2$Zn$_{20}$ and R(Fe$_x$Co$_{1-x}$)$_2$Zn$_{20}$ (R = Gd, Y) were grown
from high temperature solutions \cite{can92a} rich in Zn using initial compositions of R$_2$T$_4$Zn$_{94}$.  The
constituent elements were placed in an alumina crucible, sealed in a quartz ampule under $\sim 1/3$ atmosphere Ar
and heated in a box furnace to 1000$^\circ$ C and then slowly cooled to 600$^\circ$ C over up to 100 hours.  The
resulting single crystals were large and often manifested clear [111] facets (see the inset to figure 2a below).
For R(Fe$_x$Co$_{1-x}$)$_2$Zn$_{20}$, $x$ values are nominal values, but these are confirmed by elemental analysis
as well as compliance with Vegard's law, with the lattice parameter varying linearly between the $x = 0$ and $x =
1$ end points. Field and temperature dependent magnetization measurements were made using Quantum Design MPMS
units whereas transport and specific heat measurements were made using Quantum Design PPMS units with $^3$He
options. The electronic band structure was calculated using the atomic sphere approximation tight binding linear
muffin-tin orbital (TB-LMTO-ASA) method \cite{and75a,and84a} within the local density approximation (LDA) with
Barth-Hedin \cite{bar72a} exchange-correlation using the experimental values of the lattice parameters. The number
of atoms in the reduced unit cell is 46.  A mesh of 16 $\vec{k}$ points in the irreducible part of the Brillouin
zone (BZ) was used. The $4f$-electrons of Gd and Lu atoms were treated as a core states (polarized in the case of
Gd atoms).

Figures 1 and 2 present the temperature dependent electrical resistivity, specific heat and low field
magnetization data, as well as anisotropic $M(H)$ data,  for GdFe$_2$Zn$_{20}$ and GdCo$_2$Zn$_{20}$.  There are
two conspicuous differences between the physical properties of these compounds:  (i) GdFe$_2$Zn$_{20}$ orders
ferromagnetically whereas GdCo$_2$Zn$_{20}$ orders antiferromagnetically, and (ii) GeFe$_2$Zn$_{20}$ orders at a
remarkably high temperature of $T_C = 86$ K whereas GdCo$_2$Zn$_{20}$ orders at the more representative $T_N =
5.7$ K.   From figure 2a the high temperature Curie constant can be determined, giving effective moments ($8.05
\mu_B$ and $8.15 \mu_B$ for T=Fe and T=Co respectively) consistent with the effective moment of Gd$^{3+}$,
indicating that, in the paramagnetic state, there is little or no contribution from the transition metal.  The
saturated moment deduced from the data in figure 2b is close to that associated with Gd$^{3+}$; slightly lower for
GdFe$_2$Zn$_{20}$ and slightly higher for GdCo$_2$Zn$_{20}$. The magnetic entropy associated with each phase
transitions is approximately $R \ln 8$, with more uncertainty associated with the subtraction of the
YT$_2$Zn$_{20}$ data for T = Fe than for T = Co due to much higher ordering temperature. The remarkably high
ordering temperature found for GdFe$_2$Zn$_{20}$ is not unique to the R = Gd member of the RFe$_2$Zn$_{20}$
series. For R = Gd -– Tm transitions to ferromagnetically ordered states occur at temperatures that roughly scale
with the de Gennes parameter.  \cite{jai06a}

In order to better understand this conspicuous difference in ordering temperatures, bandstructural calculations
were carried out.  Figure 3 presents the density of states as a function of energy for both LuFe$_2$Zn$_{20}$ and
LuCo$_2$Zn$_{20}$. The upper curve in each panel is the total density of states whereas the lower curve is the
density of states associate with the transition metal ion.  It should be noted that the difference between
LuFe$_2$Zn$_{20}$ and LuCo$_2$Zn$_{20}$ density of states can be rationalized in terms of the rigid band
approximation, with the Fermi level for LuCo$_2$Zn$_{20}$ being $\sim 0.3$ eV higher than that for
LuFe$_2$Zn$_{20}$, associated with the two extra electrons per formula unit. Calculations done on YFe$_2$Zn$_{20}$
and GdFe$_2$Zn$_{20}$ as well as on YCo$_2$Zn$_{20}$ and GdCo$_2$Zn$_{20}$ lead to similar density of states
curves \cite{jai06a} and further analysis of the GdFe$_2$Zn$_{20}$ and GdCo$_2$Zn$_{20}$ banstructural results
leads to the prediction that for GdFe$_2$Zn$_{20}$ the ground state will be ferromagnetic with a total saturated
moment of approximately 6.8 $\mu_B$ (with a small induced moment on the Fe opposing the Gd moment) and for
GdCo$_2$Zn$_{20}$ the ground state will be antiferromagnetic with a saturated moment of 7.15 $\mu_B$ (and
practically no induced moment on Co). These results are consistent with the saturated values of the magnetization
seen in Fig. 2b.

These calculations indicate that the RFe$_2$Zn$_{20}$ compounds should manifest a higher $N(E_F)$ than the
RCo$_2$Zn$_{20}$ analogues and bring up the question of whether or not this is the primary reason for the
remarkably high $T_C$ found for GdFe$_2$Zn$_{20}$. In addition they bring up the question of how correlated the
electronic state is in the nominally non-magnetic Lu- or Y- based analogues.  In order to address these questions
two substitutional series were grown: Y(Fe$_x$Co$_{1-x}$)$_2$Zn$_{20}$ and Gd(Fe$_x$Co$_{1-x}$)$_2$Zn$_{20}$.
Figure 4 presents thermodynamic data taken on the Y(Fe$_x$Co$_{1-x}$)$_2$Zn$_{20}$ series.  For $x = 0$ the low
temperature, linear component of the specific heat ($\gamma$) is relatively small (19 mJ/mol K$^2$) and the
susceptibility is weakly paramagnetic and essentially temperature independent.  As $x$ is increased there is a
monotonic (but clearly super-linear) increase in the samples' paramagnetism as well as, for larger $x$-values, an
increase in the low temperature $\gamma$ values.  For YFe$_2$Zn$_{20}$ ($x = 1$) the value of gamma has increased
to over 250\% of that for YCo$_2$Zn$_{20}$ and the susceptibility has become both large and temperature dependent.
Figure 5 presents data on the analogous Gd(Fe$_x$Co$_{1-x}$)$_2$Zn$_{20}$ series.  As $x$ is increased the ground
state rapidly becomes ferromagnetic and the transition temperature increases monotonically (but again in a
super-linear fashion) and the high field, saturated, magnetization decreases weakly, in a monotonic fashion.

Taken together, figures 4 and 5 demonstrate a clear correlation between $x$, the linear component of the
electronic specific heat, the enhanced magnetic susceptibility of the Y-based series and the Curie temperature and
the saturated magnetization of the Gd-based series.  This correlation can be more clearly seen if the relation
between the linear component of the specific heat and the low temperature susceptibility of the Y-based series is
placed in the context of a nearly ferromagnetic Fermi liquid:  i.e. if the Stoner enhancement parameter, $Z$, for
each member of the series can be determined.\cite{zim72a}  For such systems the static susceptibility (corrected
for the core diamagnetism \cite{cd}) is $\chi = \chi_0/(1-Z)$, where $\chi_0 = \mu_B N(E_F)$ is the Pauli
paramagnetism. Given that the linear component of the specific heat is given by $\gamma_0 = (\pi k_B)^2 N(E_F)/3$,
if both the low temperature specific heat and magnetic susceptibility can be measured, then the parameter $Z$ can
be deduced, ($Z = 1 - (3\mu_0^2/\pi k_B^2)(\gamma_0/\chi_0)$). The canonical example of such a system is elemental
Pd for which, using data from ref. \onlinecite{cho68a},  $Z = 0.83$. For YFe$_2$Zn$_{20}$, $Z = 0.89$, a value
that places it even closer to the Stoner limit that Pd. It should be noted that the temperature dependent
susceptibility of YFe$_2$Zn$_{20}$ is remarkably similar to that of Pd as well (see ref. \onlinecite{zel04a} and
references therein). The $x$-dependence of the experimentally determined values of $\gamma$ and $\chi(T = 0)$, as
well as the inferred value of $Z$, for the Y(Fe$_x$Co$_{1-x}$)$_2$Zn$_{20}$ series are plotted in Fig. 6a. By
choosing $x$, Y(Fe$_x$Co$_{1-x}$)$_2$Zn$_{20}$ can be tuned from being exceptionally close to the Stoner limit to
well removed from it. Corrections to these inferred $Z$ values coming from the difference between the measured
electronic specific heat coefficient, $\gamma$, and Sommerfeld coefficient, $\gamma_0$, where $\gamma = \gamma_0
(1+\lambda)$ only serve to slightly increase $Z$ since $\lambda$, the electron mass enhancement parameter, is
positive definitive. We can estimate $\lambda = 0.85$ and 0.22, for $x = 1$ and $x=0$ respectively, and this
shifts $Z$ to 0.94 for YFe$_2$Zn$_{20}$ and to 0.50 for YCo$_2$Zn$_{20}$.

When the non-magnetic Y ion is replaced by the large, Heisenberg moment associated with the $S = 7/2$ Gd$^{3+}$
ion, as $x$ is varied from zero to one in the Gd(Fe$_x$Co$_{1-x}$)$_2$Zn$_{20}$ series, the Gd local moments will
be in an increasingly polarizable matrix, one that is becoming a nearly ferromagnetic Fermi liquid.  This results
in an increasingly strong coupling between the Gd local moments as $x$ is increased.  Figure 6b plots the
$x$-dependence of $T_C$ and $\mu_{sat}$ for the Gd(Fe$_x$Co$_{1-x}$)$_2$Zn$_{20}$.  The value of $T_C$ increases
in a monotonic but highly non-linear fashion in a manner reminiscent to the behavior associated with the
increasingly polarizability of Y(Fe$_x$Co$_{1-x}$)$_2$Zn$_{20}$ seen in figure 6a. The $\mu_{sat}$ value for the
field applied along the [111] direction varies systematically from the slightly enhanced 7.3 $\mu_B$ value found
for GdCo$_2$Zn$_{20}$ to the slightly deficient value of 6.5 $\mu_B$ found for GdFe$_2$Zn$_{20}$.

In addition to $x$-dependence, this conceptually simple and compelling framework can also be used to understand
the rather curious temperature dependence of the $1/\chi$ vs. $T$ data for GdFe$_2$Zn$_{20}$ seen in Fig. 2a.   As
$T$ is decreased the electronic background that the Gd$^{3+}$ ion is immersed in becomes increasingly polarizable
(as shown in Fig. 4a for $x = 1$), leading to a temperature dependent coupling that in turn leads to the
non-linearity of the $1/\chi$ vs. $T$ data.  If a constant effective moment for the Gd$^{3+}$ is assumed, then a
temperature dependent, paramagnetic $\Theta$ can be extracted from the data in Fig. 2a:  $\chi(T) = \chi_0 + C/(T
+ \Theta)$. This temperature dependent $\Theta$, shown in Fig. 4a, tracks the electronic susceptibility of the
YFe$_2$Zn$_{20}$ remarkably well, both increasing by $\sim 1.7$ upon cooling from 300 K to 100 K.

One consequence of placing Gd ions into a matrix so close to the Stoner limit is an enhanced sensitivity to small
sample-to-sample variations.  This is most clearly illustrated by the data for the
Gd(Fe$_{0.88}$Co$_{0.12}$)$_2$Zn$_{20}$ samples shown in Figs. 5a and 6b.  Although the samples have the same
nominal composition there is a clear difference in their transition temperatures (Fig. 5a).  This difference
though is not too significant given the large $dT_C/dx$ slope seen in Fig. 6b.  On the other hand, measurements on
four separate samples of Gd(Fe$_{0.25}$Co$_{0.75}$)$_2$Zn$_{20}$ did not show any significant variations in $T_C$.
Such sensitivity of correlated electron states to small disorder is not uncommon, giving rise to significant
variation in measured $T_C$ for samples of Sc$_3$In and ZrZn$_2$ [\onlinecite{mor85a,flo06a}] as well as dramatic
changes in the transport properties of heavy fermions such as YbNi$_2$B$_2$C. \cite{avi02a}

In summary, the R(Fe$_x$Co$_{1-x}$)$_2$Zn$_{20}$ series offers a unique opportunity to systematically study the
evolution of a nearly ferromagnetic Fermi liquid (for R = Y, Lu) and also offers the possibility of studying the
effects that such a polarizable background has on the ordering of large local moments which are located on an
existing, unique crystallographic site. The broader RT$_2$Zn$_{20}$ family of compounds offers an even larger
phase space for the study of correlated electron physics (for T = Fe, Ru, and Os as well as for R = Yb and Ce)
\cite{jai06a,tor06a} and for the study of local moment physics, all in the limit of a dilute, rare earth bearing,
intermetallic series. The study of these compounds promises to be a fruitful new phase space for several years to
come.

\begin{acknowledgments}
We are indebted to the following students and magneticians: K. Dennis, N. Ni, J. Friedrich, S.A. Law, H. Ko, E.D.
Mun, A. Safa-Sefat for help in samples' growth and characterization and to J. Schmalian and B.N. Harmon for useful
discussions. Ames Laboratory is operated for the U.S. Department of Energy by Iowa State University under Contract
No. W-7405-Eng.-82. This work was supported by the Director for Energy Research, Office of Basic Energy Sciences.
\end{acknowledgments}

\clearpage

\begin{figure}
\begin{center}
\includegraphics[angle=0,width=120mm]{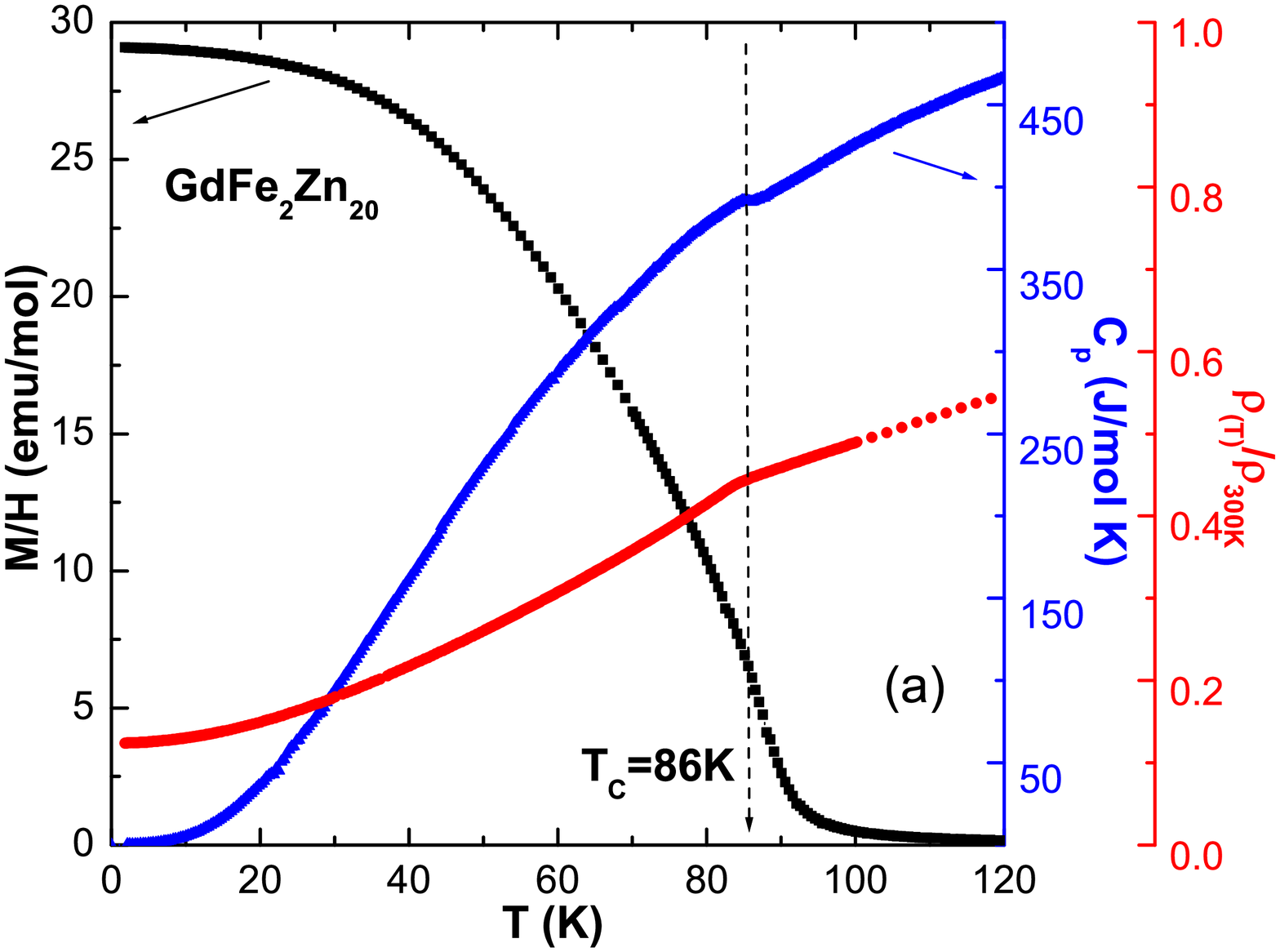}
\includegraphics[angle=0,width=120mm]{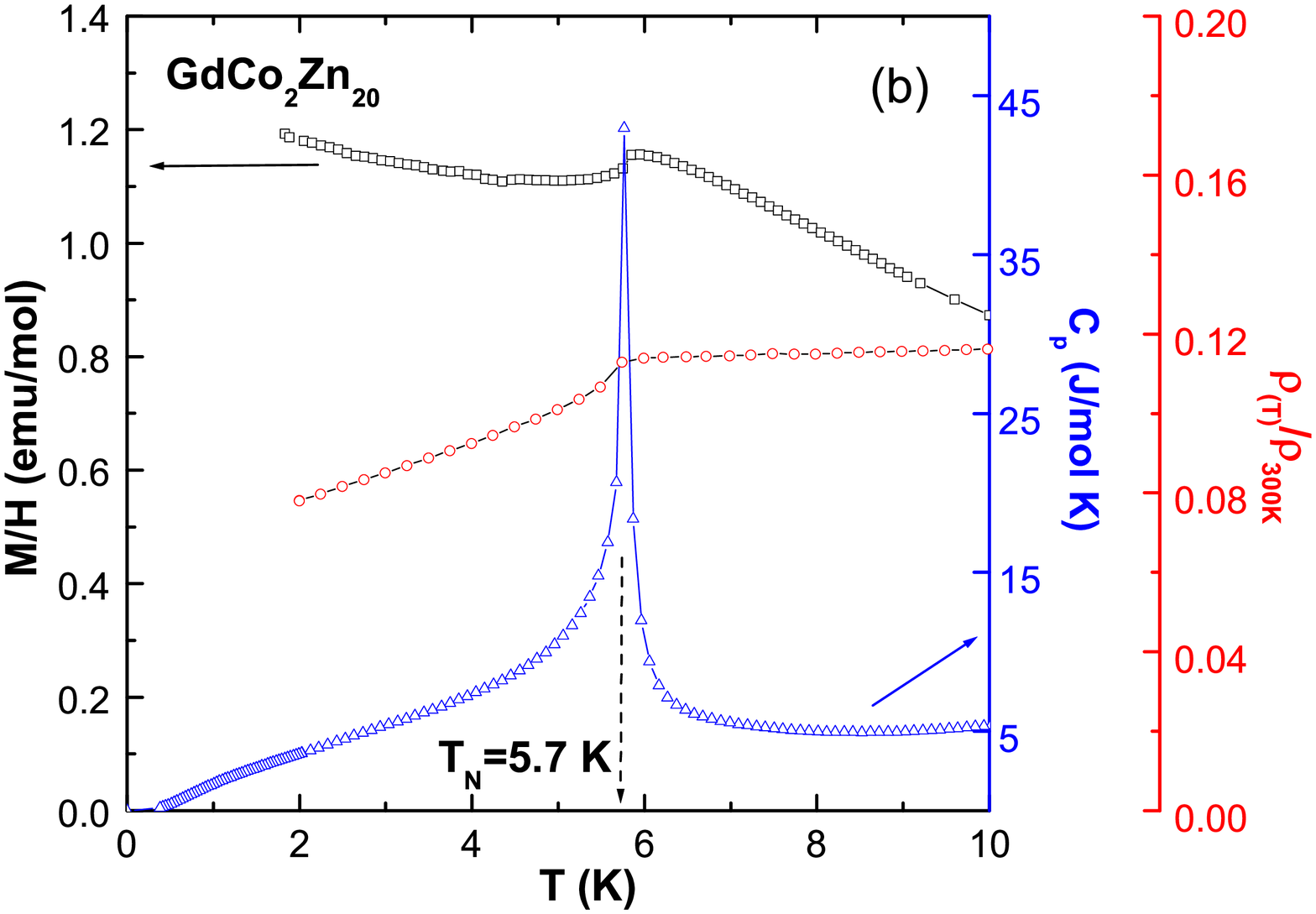}
\end{center}
\caption{Temperature dependent specific heat, resistivity, and low field ($H = 1000$ Oe) magnetization divided by
field for (a) GdFe$_2$Zn$_{20}$ and (b) GdCo$_2$Zn$_{20}$.  For GeFe$_2$Zn$_{20}$ $\rho(300 K) = 73~\mu\Omega$ cm
and for GdCo$_2$Zn$_{20}$ $\rho(300K) = 60~\mu\Omega$ cm.}\label{F1}
\end{figure}

\clearpage

\begin{figure}
\begin{center}
\includegraphics[angle=0,width=100mm]{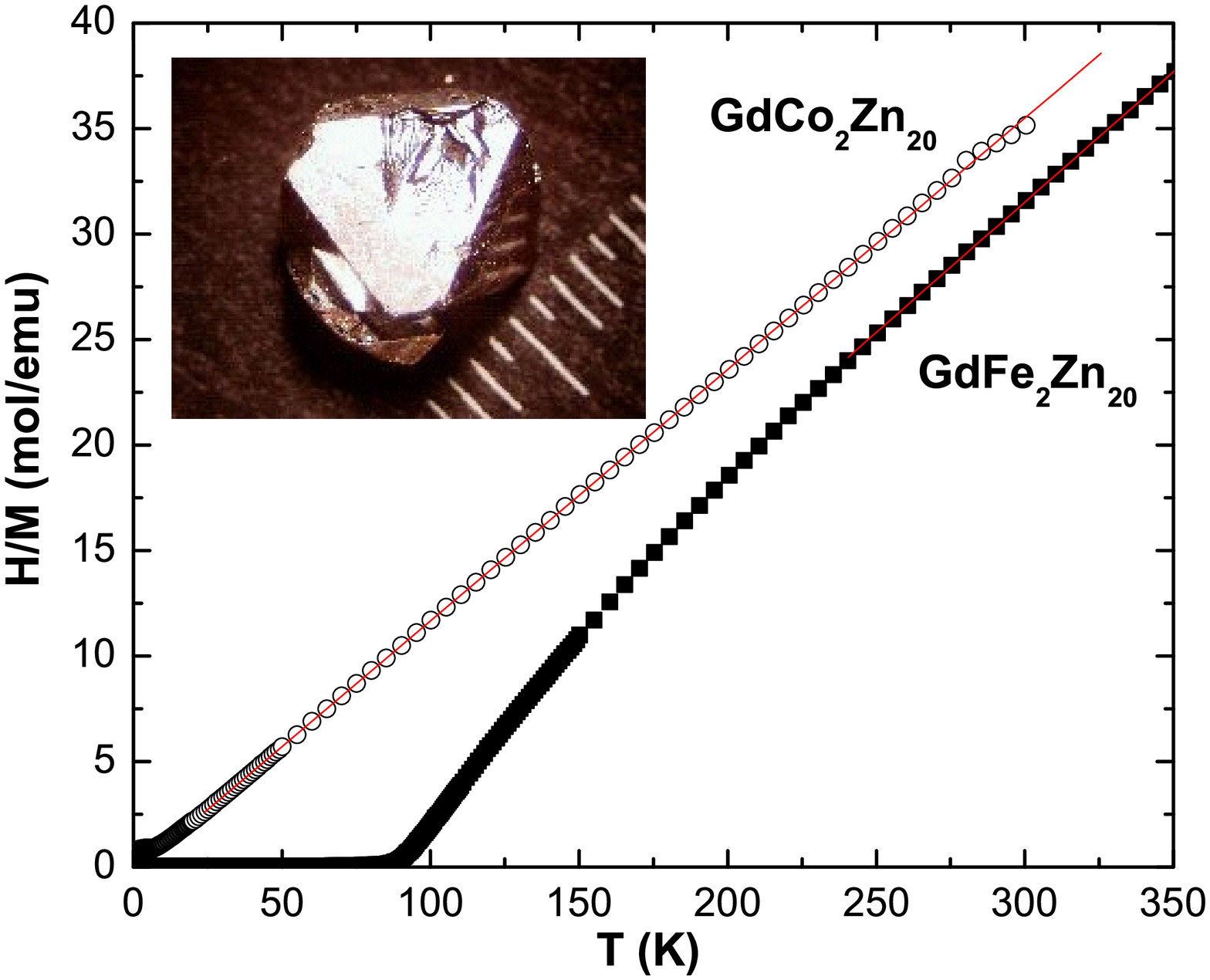}
\includegraphics[angle=0,width=100mm]{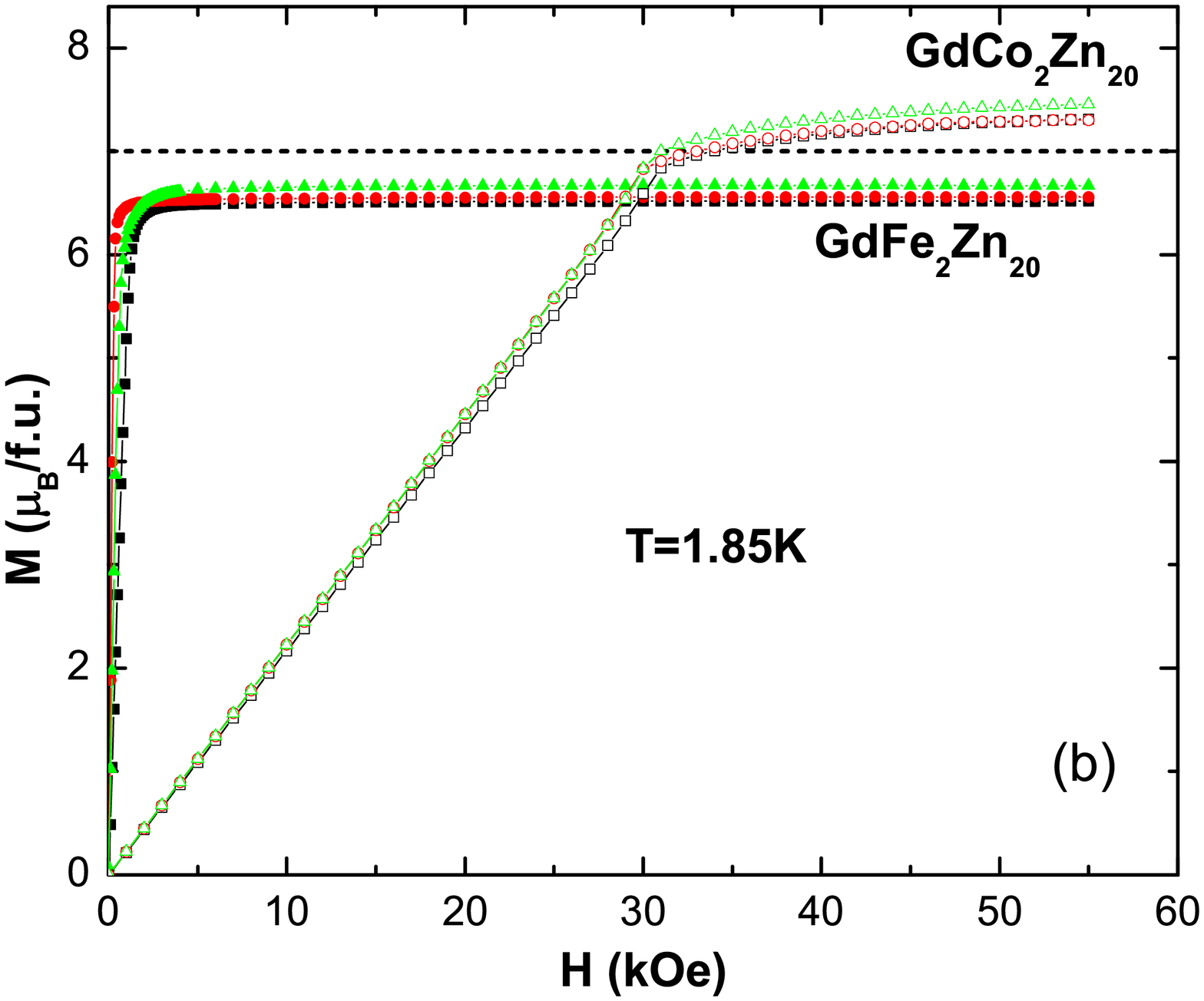}
\end{center}
\caption{(a) $H/M$ as a function of $T$ for GdFe$_2$Zn$_{20}$ and GdCo$_2$Zn$_{20}$ (dotted lines show region used
for the high temperature Curie-Weiss fit). (b) Anisotropic $M(H)$ ($T = 1.85$ K) for GdFe$_2$Zn$_{20}$ and
GdCo$_2$Zn$_{20}$. For each sample measurements for $H\|[100]$, $H\|[110]$, and $H\|[111]$ are shown. Inset to
figure 2a: a single crystal of YFe$_2$Zn$_{20}$ next to a mm scale.}\label{F2}
\end{figure}

\clearpage

\begin{figure}
\begin{center}
\includegraphics[angle=270,width=120mm]{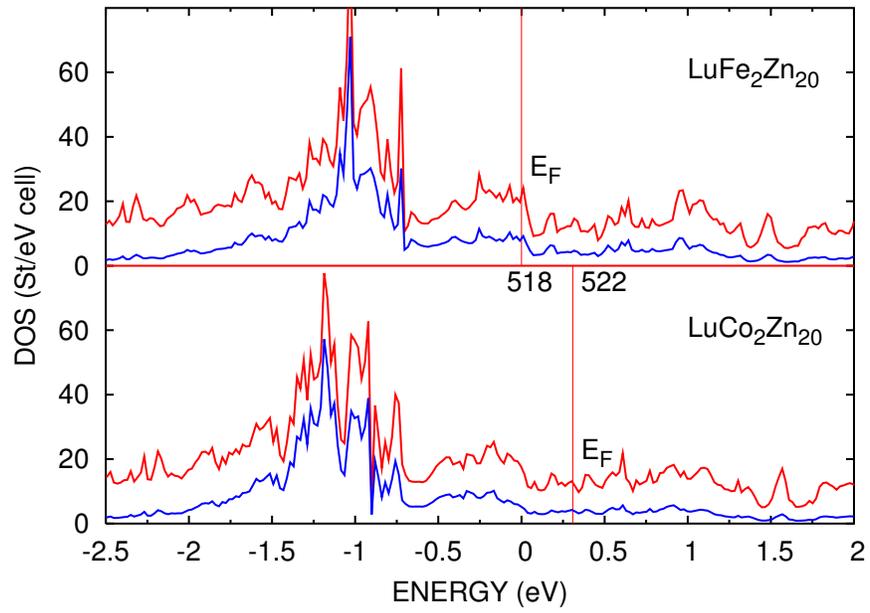}
\end{center}
\caption{Density of states as a function of energy for LuFe$_2$Zn$_{20}$ and LuCo$_2$Zn$_{20}$.  The upper curve
is total density whereas the lower curve is partial density of states associated with Fe or Co.}\label{F3}
\end{figure}

\clearpage

\begin{figure}
\begin{center}
\includegraphics[angle=0,width=120mm]{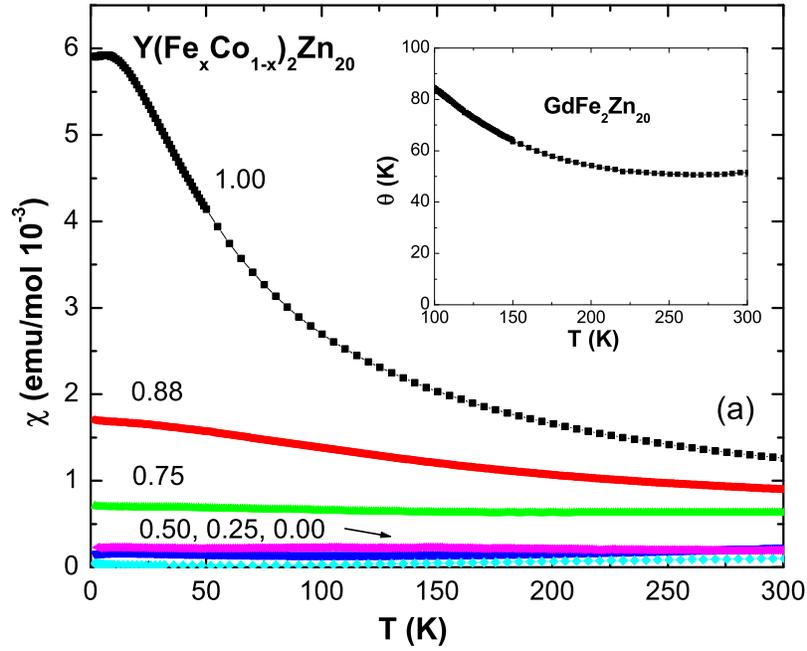}
\includegraphics[angle=0,width=120mm]{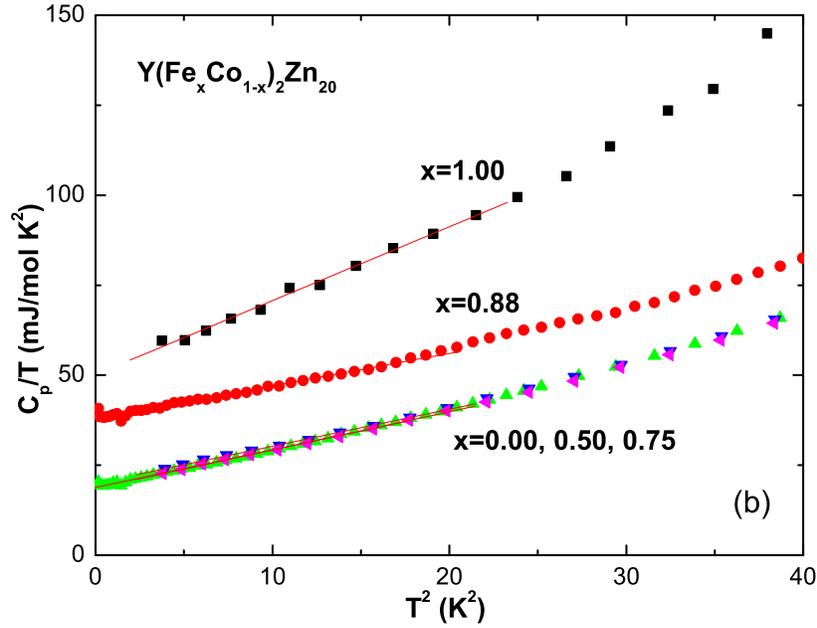}
\end{center}
\caption{Temperature dependent magnetic susceptibility (a) and low temperature $C/T$ as a function of $T^2$ (b)
for Y(Fe$_x$Co$_{1-x}$)$_2$Zn$_{20}$ series. Inset to figure 4a: temperature dependence of paramagnetic $\Theta$
for GdFe$_2$Zn$_{20}$ (see text).}\label{F4}
\end{figure}

\clearpage

\begin{figure}
\begin{center}
\includegraphics[angle=0,width=100mm]{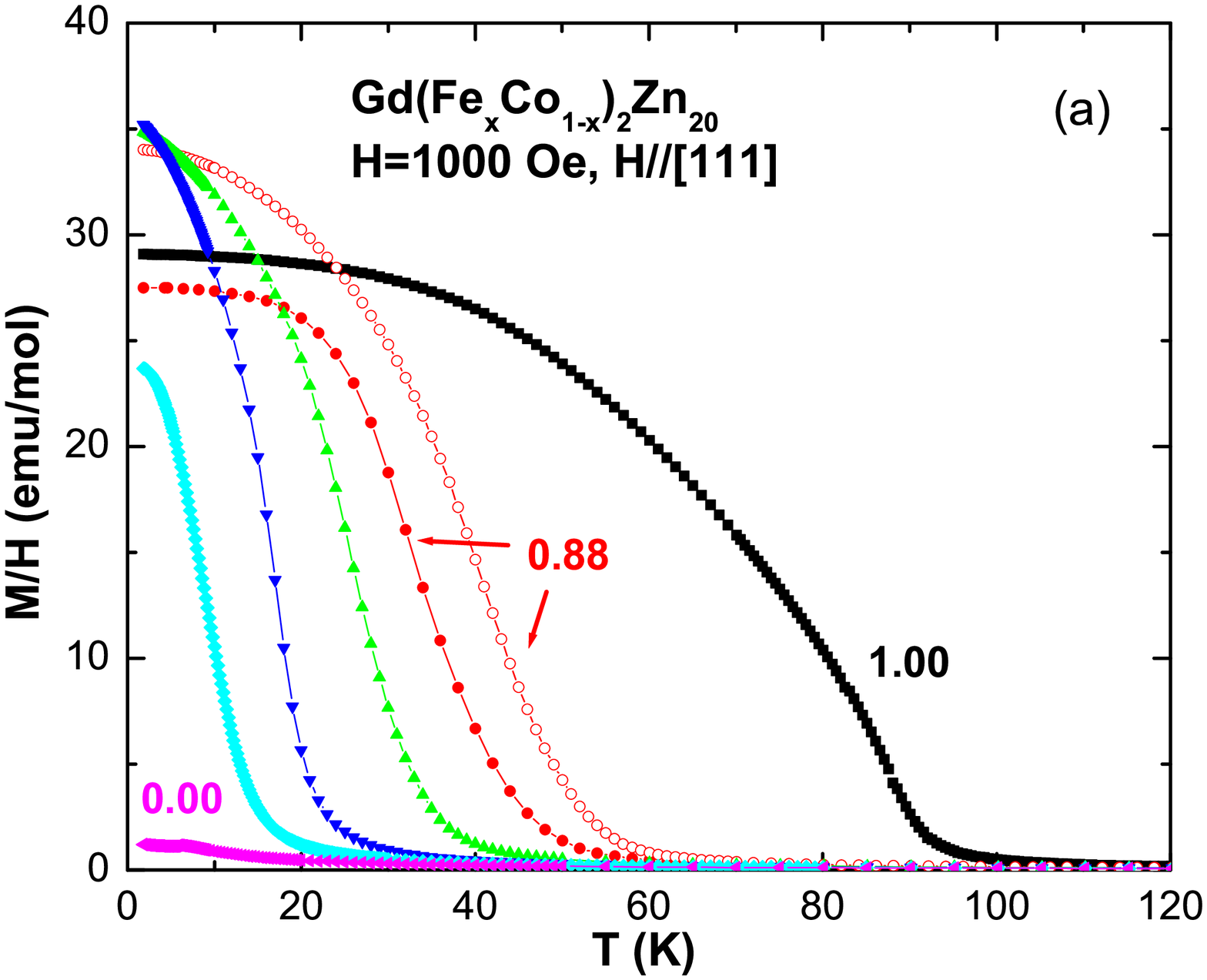}
\includegraphics[angle=0,width=100mm]{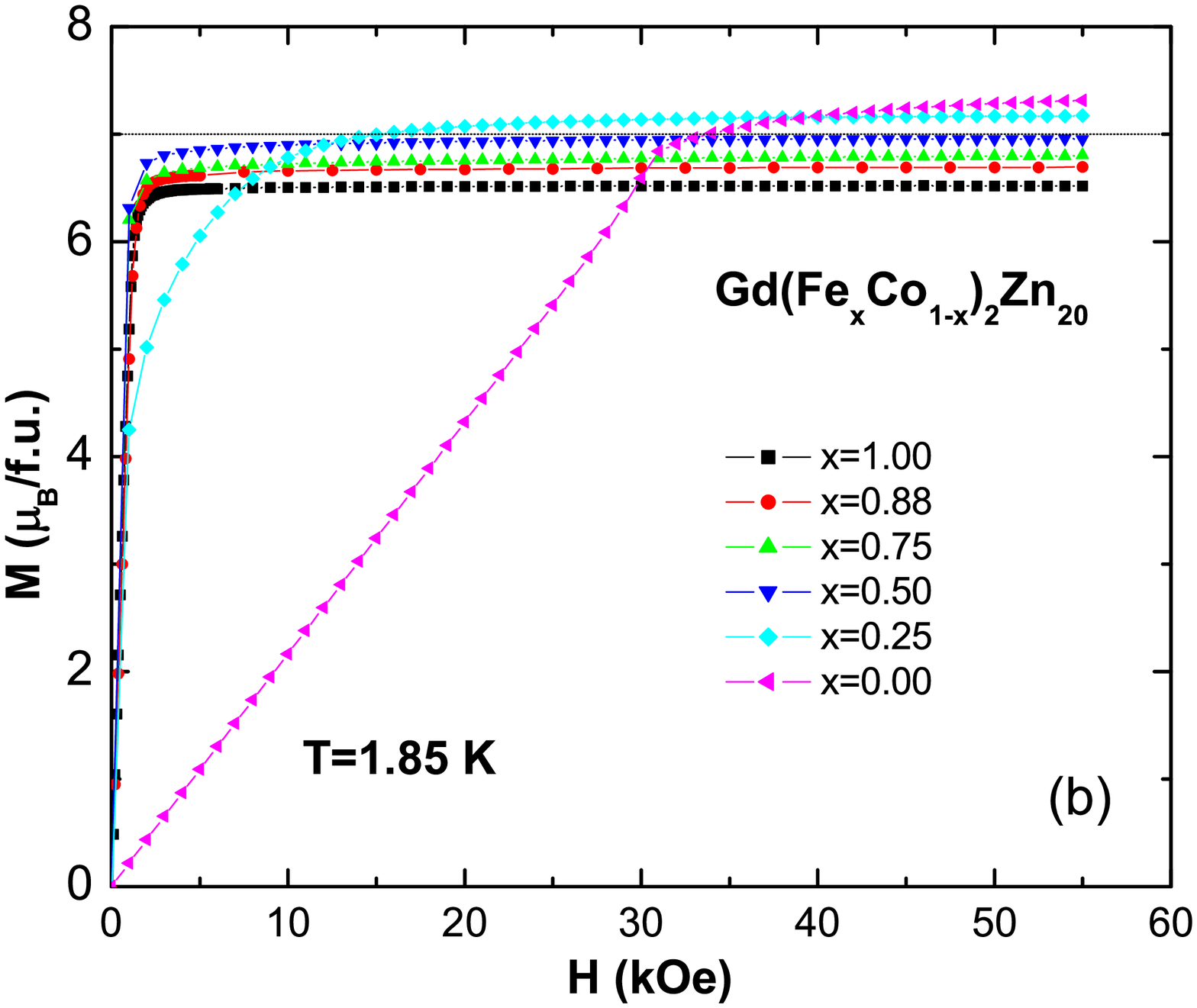}
\end{center}
\caption{Low field magnetization divided by field as a function of temperature for $x$ = 1.00, 0.88, 0.75, 0.50,
0.25, and 0.00 - from right to left (a) and low temperature ($T = 1.85$ K) magnetization as a function of applied
field (b) for Gd(Fe$_x$Co$_{1-x}$)$_2$Zn$_{20}$ series. Note that in (a) data from two samples of $x = 0.88$ are
shown. }\label{F5}
\end{figure}

\clearpage

\begin{figure}
\begin{center}
\includegraphics[angle=0,width=100mm]{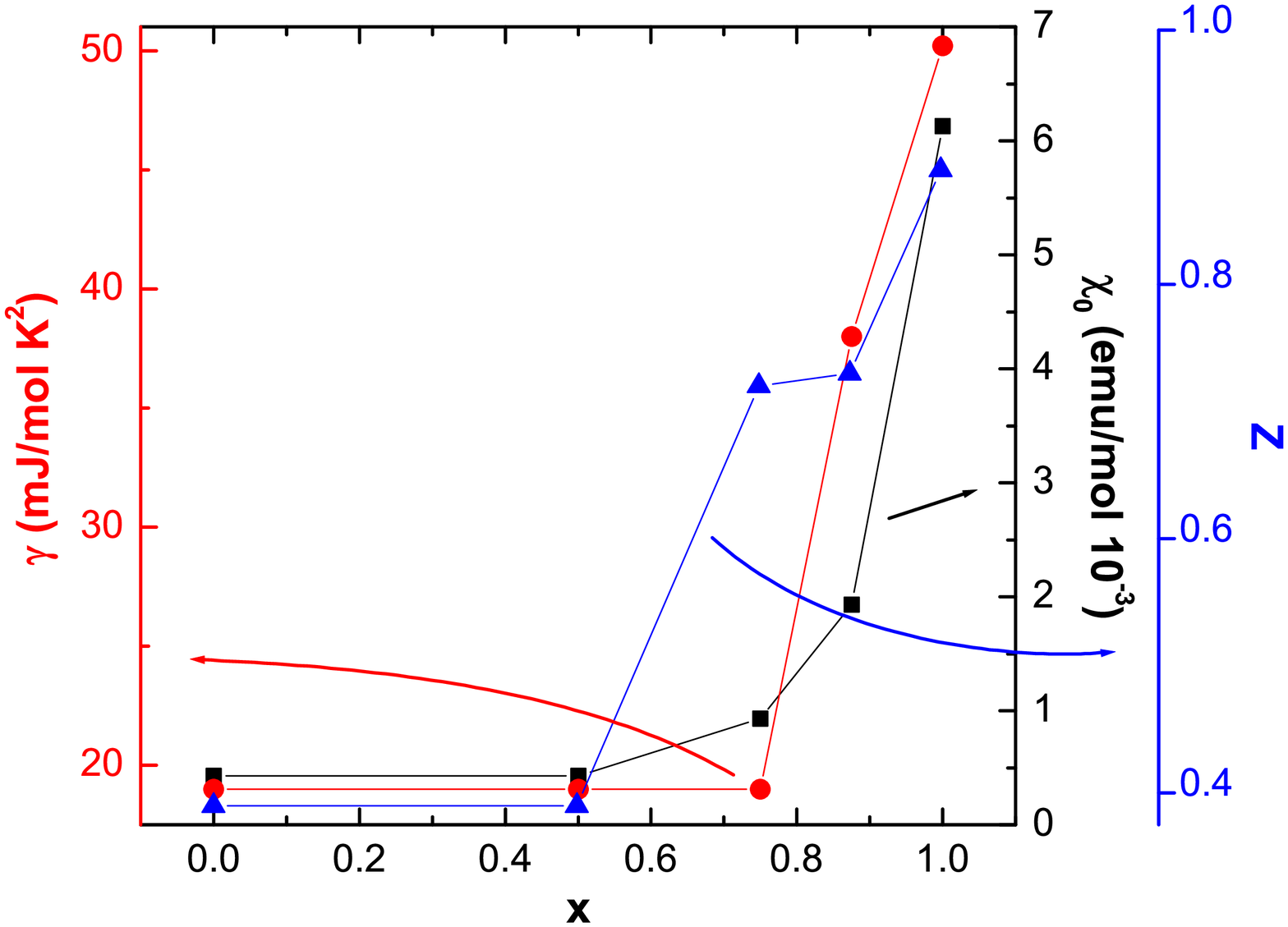}
\includegraphics[angle=0,width=100mm]{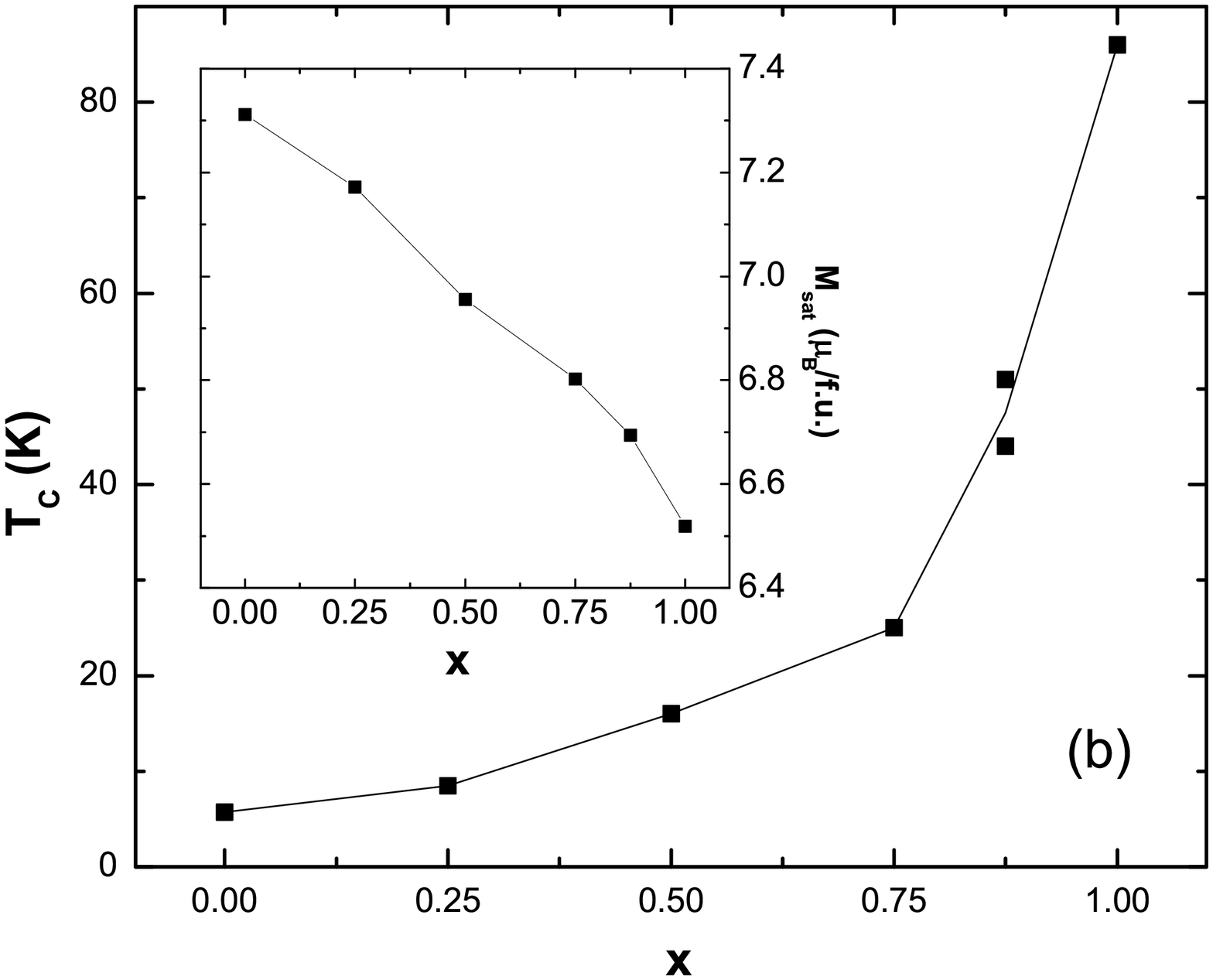}
\end{center}
\caption{Plots of linear coefficient of the specific heat, $\gamma$, magnetic susceptibility for $T \to 0$
corrected for core diamagnetism \cite{cd}, $\chi_0$,  and Stoner enhancement factor, $Z$, for
Y(Fe$_x$Co$_{1-x}$)$_2$Zn$_{20}$ series (a) and $T_C$ for Gd(Fe$_x$Co$_{1-x}$)$_2$Zn$_{20}$ series (b). Inset to
(b) shows $M_{sat}$ as a function of $x$ for Gd(Fe$_x$Co$_{1-x}$)$_2$Zn$_{20}$ series.}\label{F6}
\end{figure}

\end{document}